\documentclass[lettersize,journal]{IEEEtran}
\usepackage{amsmath,amsfonts,nccmath}
\usepackage{algorithmic}
\usepackage{algorithm}
\usepackage{array}
\usepackage[caption=false,font=normalsize,labelfont=sf,textfont=sf]{subfig}
\usepackage{textcomp}
\usepackage{stfloats}
\usepackage{url}
\usepackage{verbatim}
\usepackage{graphicx}
\usepackage{cite}
\usepackage{multirow}
\usepackage[flushleft]{threeparttable}
\usepackage{diagbox}
\usepackage{orcidlink}
\usepackage{hyperref}
\hypersetup{
	colorlinks   = true, 
	urlcolor     = black, 
	linkcolor    = black, 
	citecolor   = black 
}
\begin{document}
	
	\title{A Modal Analysis of Electromagnetic Fields Coupling into an Open-Ended Waveguide Mounted on a Finite Flange: Evaluation of a Rectangular Waveguide with a Square Flange}
	
	\author{Mohammad Eskandari\orcidlink{0009-0002-0599-2742}, Mojtaba Joodaki\orcidlink{0000-0002-2239-1271},~\IEEEmembership{Senior Member,~IEEE}, Amir Reza Attari
		\thanks{M. Eskandari is with the School of Computer Science and Engineering, Constructor University Campus Ring 1, 28759 Bremen, Germany (e-mail: meskandari@constructor.university).} 
		\thanks{M. Joodaki is with the School of Computer Science and Engineering, Constructor University, Campus Ring 1, 28759 Bremen, Germany (e-mail: mjoodaki@constructor.university).}
		\thanks{A. R. Attari is a professor of Electrical Engineering at Department of Electrical Engineering, Faculty of Engineering, Ferdowsi University of Mashhad, Mashhad, Iran (e-mail: attari50@um.ac.ir).}}
	
	\markboth{----}%
	{ESKANDARI, JOODAKI, AND ATTARI: A MODAL ANALYSIS OF ELECTROMAGNETIC FIELDS COUPLING...}
	
	\maketitle
	
	\begin{abstract}
		This paper presents a modal analysis based on the reciprocity theorem to calculate the coupled/penetrated fields into an open-ended waveguide mounted on a finite flange. Although there is no limitation on the geometry and type of the external source, an infinitesimal dipole is chosen to produce a plane wave incident to the waveguide aperture. The proposed method relates the amplitude of each penetrated mode into the waveguide to the far-field radiation components of that mode from the waveguide aperture. The accuracy of the final result depends on the accuracy of the calculated radiated fields. The radiated field components are calculated considering the reflected fields due to the aperture and the diffracted fields due to the flange edges. 
		For the first time, the impact of a finite flange on the penetrated fields into a waveguide is discussed comprehensively.
		The geometry of a rectangular waveguide mounted on a thick square flange is selected to be evaluated. The effects of changing the main parameters of the geometry on the penetrated fields are discussed rigorously in various examples. Our results are compared with 3D full-wave simulations, and an excellent agreement was found between the results while our analytical approach showing a 60 times faster performance. In addition, we compared our results with the measurement reported in a previous study.
	\end{abstract}
	
	\begin{IEEEkeywords}
		Electromagnetic fields penetration, finite flange, oblique plane-wave incidence, reciprocity theorem, rectangular waveguide analysis.
	\end{IEEEkeywords}
	
	\section{Introduction}
	\IEEEPARstart{E}{lectromagnetic} penetration through apertures on a conducting screen has been one of the most interesting classical problems in electromagnetic compatibility (EMC) and microwave engineering \cite{ref1,ref2,ref3,ref4,ref5,ref6,ref7,ref8,ref9,ref10,ref11,ref12,ref13}. This analysis can be directly used in various fields such as antenna design, radar, and satellite communication. It can also play an important role in electromagnetic shielding design,  waveguide filters design, and sensor development. One of the first theories on the topic of coupling through aperture was proposed by Bethe \cite{ref1} for an small circular aperture on conducting surface using the concept of equivalent magnetic current. There was an error in determination of magnetic current that was corrected in \cite{ref2}. Numerous studies have used various methods to solve different cases of the problem of penetration through apertures, and more than 100 of the earlier studies are referenced in \cite{ref3}. The generalized network approach presented in \cite{ref4} was the basis of many studies on the subject of coupling between aperture on an infinite conducting plane that mentioned in \cite{ref5}. A method based on Kobayashi Potentials (KP) is presented in \cite{ref6} to solve the problem of diffraction by a rectangular hole in a conducting plate. Recently, Their method has been extended to solve the transmission and diffraction by a rectangular hole in a thick conducting screen \cite{ref7}. Their solution is given for two half-space and one waveguide region. An approach based on the method of moment (MoM) is presented in \cite{ref8} to evaluate a thick rectangular window in receiving mode. Most of the methods mentioned above are based on modal analysis which are efficient for low and middle range frequencies. However, at high frequencies (HF) when the dimensions are significantly large compared to the wavelength they are not computationally efficient because a huge number of modes must be considered in the solution. Thus, some hybrid and selective modal methods in \cite{ref9,ref10,ref11,ref12} are presented to solve open-ended waveguides and cavities at HF. Recently, a novel spectral method proposed in \cite{ref13} that models arbitrarily flanged dielectric-loaded parallel plate waveguide semi-analytically using the generalized Wiener–Hopf technique (GWHT) focusing on both diffraction phenomena and guided waves through the continuous and discrete spectra.
	
	Several methods have been proposed in \cite{ref14} to calculate diffracted fields due to the straight and curved edges. Some of these methods are utilized in \cite{ref15} to evaluate the impact of the diffracted fields due to the edges of the ground plane (flange) on the radiated fields from the aperture antennas. The results of \cite{ref15} are calculated only for the dominant mode of the waveguide without considering the effect of reflection due to the aperture. In \cite{ref16}, we proposed an analytical approach to calculate the reflected fields due to the rectangular waveguide aperture for any incident mode. We use this approach to obtain highly accurate results of the radiated field components using Geometrical Theory of Diffraction (GTD) introduced in \cite{ref17}. However, GTD is most effective for high-frequency waves when the wavelength is much smaller than the dimensions of the scattering object. This theory is applicable when interacting with large objects with sharp edges in a linear, homogeneous, and isotropic (LHI) medium. The accuracy of GTD diminishes in the near-field region, especially when multiple scattering occurs or there is significant surface roughness.
	
	Analytical methods are highly regarded in various scientific and engineering fields because they provide quick and straightforward solutions, as well as deep insights into the underlying principles of the problems being studied. Unlike experimental and numerical methods, which frequently require substantial time investments (often spanning several hours or more), analytical techniques usually offer a swifter and more cost-effective alternative enabling researchers to predict the system behavior without the need for extensive simulations. Furthermore, analytical solutions present the system behavior in a mathematical expression that allows investigating the impacts of several parameters on the results simultaneously with a low computational cost. Therefore, they can be conveniently implemented for system optimization purposes.
	
	In this paper, we present a modal analysis of electromagnetic (EM) field coupling/penetration into an open-ended waveguide mounted on a finite flange. Our goal in this paper is to demonstrate the effect of finite flange on the penetrated fields into the waveguide for the first time. Achieving this goal could open a new way for accurate analyzing the shielding effectiveness (SE) of a metal enclosure with apertures \cite{ref18,ref19,ref20,ref21}. Also, another major aim of this paper is to obtain accurate results of the radiated fields from the rectangular waveguides by considering the effects of both diffracted fields from the aperture and the flange edges. It was stated in \cite{ref8} that the reciprocity cannot be applied straightway to the problem of penetration into a waveguide because of the different types of excitation between transmit and receive modes. However, we could solve this problem by applying the reciprocity theorem for each waveguide mode separately to relate the penetrated fields into the waveguide to its accurate radiated fields from the waveguide. Our solution is obtained analytically, where the accuracy of our final modal results is influenced by two primary factors. Firstly, the number of modes considered in calculating aperture reflected fields, where the corresponding errors can be minimized using the algorithm detailed in \cite{ref16}. Secondly, the inherently approximate nature of diffracted field calculations, as discussed in \cite{ref14}. Thus, our method can calculate the coupled fields into the waveguide mounted on an infinite flange with high accuracy. However, we show that this is not sufficient to achieve accurate results in practical cases. We examine the geometry of a rectangular waveguide mounted on a thick square flange. We investigate this problem from different aspects such as: operating frequency, flange size, plane wave incidence angle and polarization direction, and the waveguide dimensions. Also, the results of higher-order modes penetration are shown. All of our analytical results are validated against the accurate results of 3D full-wave simulations, and in one case, they are also validated against the measurement results presented in \cite{ref8}. At the end of this paper, we present a comparison between the runtime of our analytical approach program code and the 3D full-wave simulator, and the efficiency of our approach will be discussed.

	\section{Analysis}
	Fig. 1 shows an open-ended waveguide along the z-axis with an arbitrary cross-section mounted on a finite arbitrary-shaped perfect electric conductor (PEC) flange.
	\begin{figure}[!t]
		\centering
		\includegraphics[width=3.4in]{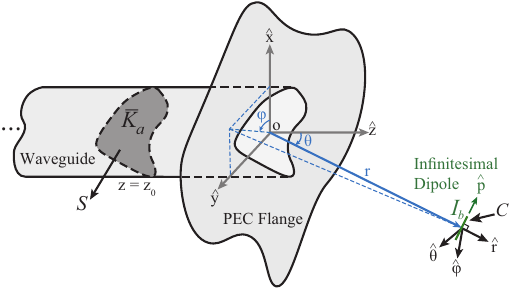}
		\caption{An open-ended waveguide with an arbitrary cross-section mounted on a finite, arbitrarily shaped PEC flange radiating into free space by a surface current of $K_a$ on the waveguide cross-section surface $S$ at $z=z_0$, and an infinitesimal dipole with a current of $I_b$ in the direction of $\hat{p}$ on the path $C$, placed far from the waveguide aperture at a distance $r$.}
	\end{figure}
	The spherical and cartesian coordinates are shown, and the coordination origin is placed at the center of the waveguide aperture. Throughout this paper, superscripts $h$ and $e$ denote TE and TM modes, respectively. Also, time-harmonic dependence $e^{j{\omega}t}$ is assumed and omitted from the calculations.
	
	\subsection{The Waveguide Modes Fields Distribution}
	It is evident that the waveguide in Fig. 1 only supports TE and TM modes. Let $z = z_0$ be the zero phase reference plane. Then, the transverse electric and magnetic fields of each waveguide mode traveling in $+z$ direction can be generally written as
	\begin{subequations}
		\begin{equation}
			\bar{E}^+_{t,m} = [E_{x,m}(x,y) \hat{x} + E_{y,m}(x,y) \hat{y}]e^{-\gamma_{m}(z-z_0)}
		\end{equation}
		\begin{equation}
			\bar{H}^+_{t,m} = Y_{m}[-E_{y,m}(x,y) \hat{x} + E_{x,m}(x,y) \hat{y}]e^{-\gamma_{m}(z-z_0)}
		\end{equation}
	\end{subequations}
	where $m$ denotes the waveguide mode index, $E_{x,m}$ and $E_{y,m}$ are the scalar functions related to the electric fields of the mode, $\hat{x}$ and $\hat{y}$ are the unit vector along the $x$ and $y$ axes, respectively, $\gamma_{m}$ is the complex propagation constant of the mode, and $Y_{m}$ is the mode characteristic admittance. If we assume the fields given in (1) are produced by a fictitious electric surface current $\bar{K}_a$ on $S$ at $z=z_0$, then the electric field of the mode traveling in $-z$ direction produced by $\bar{K}_a$ is the same as in (1) at $z=z_0$, and we can write
	\begin{subequations}
		\begin{equation}
			\bar{E}^-_{t,m} = [E_{x,m}(x,y) \hat{x} + E_{y,m}(x,y) \hat{y}]e^{+\gamma_{m}(z-z_0)}
		\end{equation}
		\begin{equation}
			\bar{H}^-_{t,m} = Y_{m}[E_{y,m}(x,y) \hat{x} - E_{x,m}(x,y) \hat{y}]e^{+\gamma_{m}(z-z_0)}.
		\end{equation}
	\end{subequations}
	Therefore, based on surface equivalence theorem \cite{ref14} we have
	\begin{equation}
		\bar{K}_a = \hat{z} \times [\bar{H}^+_{t,m} - \bar{H}^-_{t,m}]_{z=z_0}
		= -2 Y_{m} \bar{E}^\pm_{t,m}\arrowvert_{z=z_0}.
	\end{equation}

	\subsection{Reciprocity Theorem for the Presented Geometry}
	The reciprocity theorem is one of the most useful tools for analyzing antennas in receiving mode. Assume that two sources, $\bar{J}_a$ and $\bar{J}_b$, are within a linear and isotropic medium and are enclosed by a sphere with an extremely large radius, occupying volume $V$. These two sources radiate electric fields of $\bar{E}_a$ and $\bar{E}_b$, respectively, throughout the geometry. Then, based on the reciprocity theorem given in \cite{ref14}, we can write
	\begin{equation}
		\iiint_V (\bar{E}_b \cdot \bar{J}_a) dv^\prime = \iiint_V (\bar{E}_a \cdot \bar{J}_b) dv^\prime.
	\end{equation}
	
	In addition, Fig. 1 shows an infinitesimal dipole carrying electric current $I_b$ in the direction of unit vector $\hat{p}$. This dipole is placed on the plane normal to $\hat{r}$. The dipole length is $l$ on the path $C$. As the reciprocity theorem conditions are satisfied, we can apply (4) to the geometry of Fig. 1 as follows: 
	
	\begin{equation}
		\iint_{S} (\bar{E}_b \cdot \bar{K}_a) ds^\prime = \int_C (\bar{E}_a \cdot \hat{p})I_b dl^\prime
	\end{equation}
	where $\bar{E}_b$ is the penetrated electric field into the waveguide on $S$ due to the infinitesimal dipole, and $\bar{E}_a$ is the radiated electric field at the positions of the dipole due to $\bar{K}_a$. The infinitesimal dipole is placed far from the waveguide to produce an incident plane wave to the waveguide aperture. The electric field of the incident plane wave $\bar{E}_{inc}$ is given by \cite{ref22}
	\begin{equation}
		\bar{E}_{inc} = -E_{inc}\hat{p} = -j \eta_0 \frac{k_0 I_b l e^{-j k_0 r}}{4\pi r} \hat{p}
	\end{equation}
	where $\eta_0=120\pi \varOmega$ is the free space intrinsic impedance, $k_0 = 2\pi f \sqrt{\varepsilon_0 \mu_0}$  is the free space wavenumber, $f$ is the operating frequency, $\mu_0$ is the free-space permeability, $\varepsilon_0$ is the free-space permittivity, and $r$ is the distance between the dipole and the center of the aperture. Also, $\bar{E}_a$ can be written in the following form
	\begin{multline}
		\bar{E}_a = \bar{G}_a \frac{e^{-j k_0 r}}{r} = [g_\theta(\theta,\phi) \hat{\theta} + g_\phi(\theta,\phi) \hat{\phi}]\frac{e^{-j k_0 r}}{r}.
	\end{multline}
	where $\bar{G}_a$ might be referred as far-field antenna pattern function. However, in this context, we do not perform any pattern normalization and its unit is (V).
	
	All variables are supposed to be constant over the infinitesimal length of $l$. Thus, we can solve the right side of (5) by replacing $\bar{E}_a$ from (7) and $I_b$ from (6) as follows
	\begin{equation}
		\int_C (\bar{E}_a \cdot \hat{p})I_b dl^\prime = -j\frac{4\pi E_{inc}}{k_0 \eta_0} (\bar{G}_a \cdot \hat{p}).
	\end{equation}
	
	$\bar{E}_b$ can be written as the sum of transverse electric fields of the waveguide modes
	\begin{equation}
		\bar{E}_b = \sum_{n=1}^{\infty} C_{n}\bar{E}^-_{t,n}\arrowvert_{z=z_0}.
	\end{equation}
	where $C_{n}$ is a complex coefficient of the $n$th mode that penetrates into the waveguide. Then, based on orthogonality of the waveguide modes, we can simplify the left side of (5) using (3) and (9) as
	\begin{equation}
		\iint_{S} (\bar{E}_b \cdot \bar{K}_a) ds^\prime = -2 C_{m} Y_{m}\iint_{S} |(\bar{E}^-_{t,m}\arrowvert_{z=z_0})|^2 ds^\prime
	\end{equation}
	Therefore, according to (5), the right sides of both (8) and (10) are equal, and we can write
	\begin{equation}
		C_{m} = -j\frac{2\pi (\bar{E}_{inc} \cdot \bar{G}_a)}{k_0 \eta_0 Y_{m} \iint_{S} |(\bar{E}^-_{t,m}\arrowvert_{z=z_0})|^2 ds^\prime}.
	\end{equation}
	The integral in (11) can be readily solved for the conventional waveguides. Then, only $\bar{G}_a$ is left to be determined. The expression in (7) states that $\bar{G}_a$ can be obtained directly from $\bar{E}_a$. Thus, the accuracy of the penetrated fields into the waveguide is directly dependent on the accuracy of the calculated radiated fields from the waveguide.

	\subsection{Far-Field Radiation From the Aperture}
	The total electric fields radiated by $\bar{K}_a$ can be written in the following form \cite{ref15},
	\begin{equation}
		\bar{E}_a = \bar{E}^u + \bar{E}^d
	\end{equation}
	where $\bar{E}^u$ is the total electric field radiated by $\bar{K}_a$ when the waveguide is mounted on an infinite (unlimited) flange, and $\bar{E}^d$ is the diffracted fields radiated by the flange edges.
	
	Based on the equivalent current theorem \cite{ref22}, $\bar{E}^u$ can be calculated from the equivalent surface magnetic current $\bar{M}$ over the aperture surface given as
	\begin{equation}
		\bar{M} = -2\hat{z} \times \bar{E}^{ap}
	\end{equation}
	\begin{equation}
		\bar{E}^{ap} = \bar{E}^{ap,i} + \bar{E}^{ap,r}
	\end{equation}
	where $\bar{E}^{ap}$ is the total electric field over the aperture due to $\bar{K}_a$, $\bar{E}^{ap,i}$ is the incident electric field to the aperture given in (1a), and $\bar{E}^{ap,r}$ is the reflected fields from the aperture due to $\bar{E}^{ap,i}$. A comprehensive modal analysis to calculate $\bar{E}^{ap,r}$ for the rectangular waveguides is presented in \cite{ref16}. Thus, $\bar{E}^u$ for a rectangular waveguide can be calculated (see Appendix A).
	
	The diffracted fields from the flange edges can be calculated using GTD given in \cite{ref14} for the first-order diffraction as
	\begin{equation}
		\bar{E}^d(s) = \bar{E}^i(Q_d) \cdot \bar{\bar{D}}\textrm{ } A(s',s)\textrm{ } e^{-jk_0s}
	\end{equation}
	where $\bar{E}^i(Q_d)$ is the incident electric fields to the diffraction point $Q_d$ at the edge of the flange, $\bar{\bar{D}}$ is a dyadic function of the diffraction coefficient, $A(s',s)$ is the spatial attenuation factor, and $s$ is the distance from $Q_d$ to the observation point. Equation (15) can be simplified by having more information on the flange geometry, incident wavefront, and observation point distance. Also, the second-order (slope) diffraction and equivalent current contribution (see Appendix B) discussed in \cite{ref14} can be used to improve the accuracy of the results.

	\section{Numerical Results and Simulations}
	In this section, we evaluate the geometry of an open-ended rectangular waveguide mounted on a thick conducting square flange, as shown in Fig. 2.
	\begin{figure}[!t]
		\centering
		\includegraphics[width=3in]{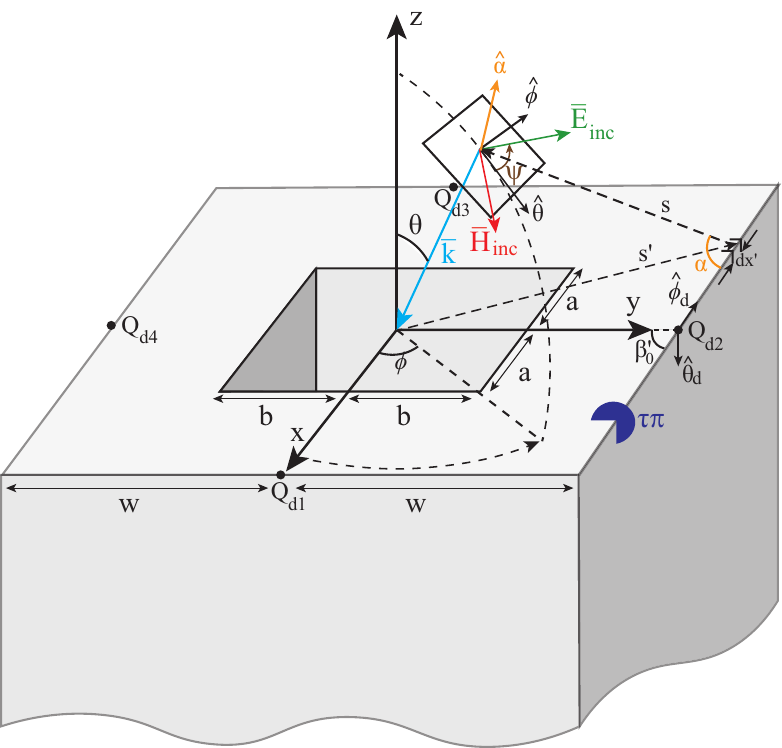}
		\caption{Geometry of an open-ended rectangular waveguide mounted on a thick square PEC flange and an external plane wave incident to its aperture.}
	\end{figure}
	The waveguide dimensions are $2a\times 2b$ in $x$ and $y$ directions, respectively. The length of the sides of the flange is $2w$, and the exterior angle of the flange edges is $\tau\pi$ radians. For the case of a thick flange, $\tau=3/2$. The medium used on both sides of the aperture is set to be free space. The cartesian and spherical coordinates are illustrated in Fig. 2. $\psi$ is the angle between $\hat{\theta}$ and the electric field of the plane wave with the wave vector of $\bar{k} = -k_0\hat{r}$. Therefore, we can write
	\begin{equation}
		\bar{E}_{inc} = E_{inc}[(cos\psi) \hat{\theta} + (sin\psi) \hat{\phi}]
	\end{equation}
	
	According to Appendix A and Appendix B, we can calculate $\bar{E}_a$ for the geometry of Fig. 2. Thus, using (16) and (7), we can simplify (11) and define normalized penetration ratio for different modes with $uv$ indices as

	\begin{equation}
		\textrm{NPR}_{uv} = \frac{A^p_{uv}}{E_{inc}} = \frac{2\pi k_0^2 (g_\theta . cos\psi + g_\phi . sin\psi)}{A^s_{uv} \gamma_{uv} ab (a_u^2 + b_v^2)\epsilon_{(u.v)}} 
	\end{equation}
	where $A^s_{uv}$ is the amplitude coefficient of the mode at the reference plane due to $\bar{K}_a$, and $A^p_{uv} = C_{uv} A^s_{uv}$ is the amplitude coefficient of the penetrated mode into the waveguide which is also at the reference plane. It should be noted that (17) is valid for both transverse electric (TE) and transverse magnetic (TM) modes of the rectangular waveguide. The reason for defining NPR is due to the limitations of existing approaches in presenting penetrated fields. They often require a unique far-field diffraction pattern for each incident direction (e.g., those presented in \cite{ref6}), whereas NPR provides a consistent interpretation similar to an antenna pattern for penetrated modes. Unlike antenna pattern normalization which highlights relative differences, NPR focusing on the ratio of penetrated to incident field amplitude which is essential for applications like shielding effectiveness. Moreover, the calculation of the Antenna Factor (AF), which will be discussed later, is limited to the dominant mode bandwidth and lacks phase information, whereas NPR offers both magnitude and phase details for penetrated modes without frequency restrictions. Since, to the best of our knowledge, there is no parameter that meets the requirements of this paper, we have defined NPR.
	
	The geometry of Fig. 2 is evaluated in the following examples for the primary values of $f = 10$ GHz, $a = 1.143$ cm, $b = 0.508$ cm, $w = 20b = 10.16$ cm, $\theta = 0$, $\phi = \pi/2$, and $\psi = 0$. These specifications are used in all subsequent examples unless otherwise specified. We evaluate the effects of changing these parameters separately on the results of the penetrated fields when $z_0 = 0$. The analytical results are computed using MATLAB 2023b, incorporating our previous work discussed in \cite{ref16} satisfying the condition set $A$ for all examples. The results from 3D full-wave simulations were obtained using CST Studio Suite 2023, based on the model illustrated in Fig. 3 where all dimensions are the same as those given above, with the flange thickness set to 3 cm. To achieve higher accuracy, the frequency domain solver is chosen for its superior performance in narrow-band applications.
	\begin{figure}[!t]
		\centering
		\includegraphics[width=3in]{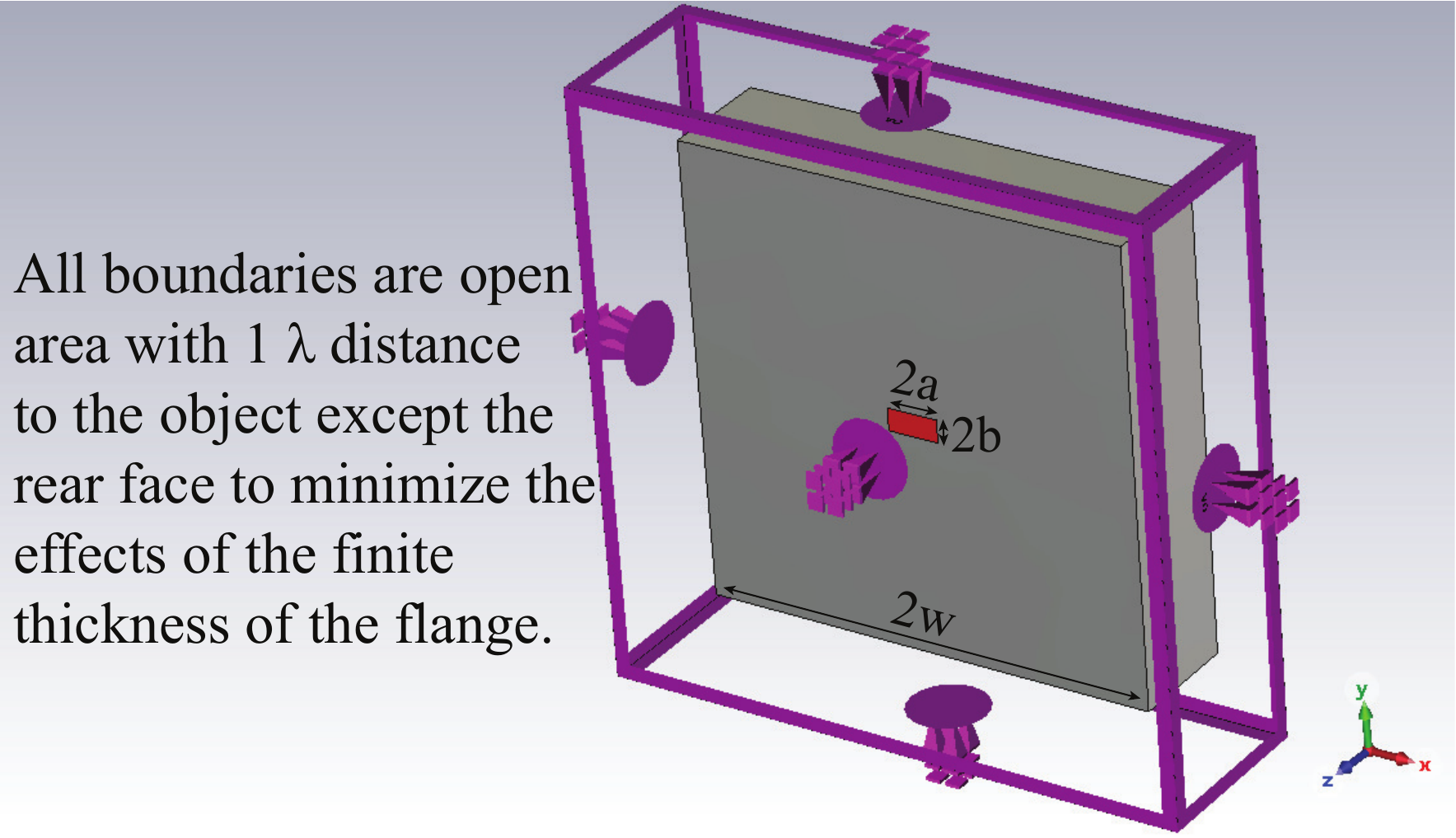}
		\caption{Geometry and boundary conditions of the implemented CST model.}
	\end{figure}
	Then, the values of NPR can be obtained directly from F-parameters calculated by the simulator by defining a waveguide port within the waveguide region and calibrating the results to the reference plane of $z=0$.

	\subsection{The Evaluation of NPR$^h_{10}$ in Terms of Frequency Changes}
	Fig. 4 shows the magnitudes and phases of NPR$^h_{10}$ over frequencies between 1 to 16 GHz.
	\begin{figure}[!t]
		\centering
		\includegraphics[width=3.49in,height=3.7in]{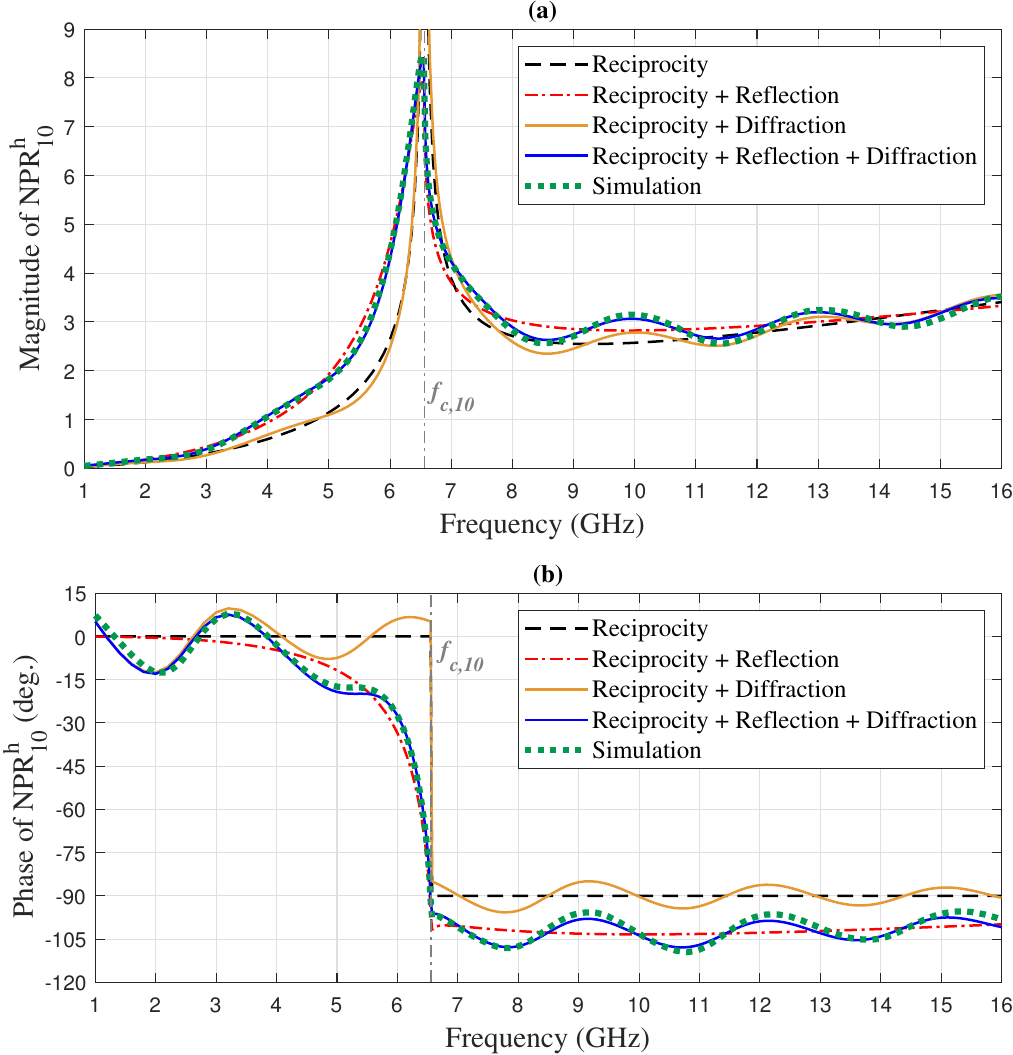}
		\caption{(a) Magnitude and (b) Phase of NPR$^h_{10}$ in terms of frequency changes.}
	\end{figure}
	The values of NPR$^h_{10}$ are calculated in four ways to show the effect of each included field on the results. It can be seen in Fig. 4 that the results of the blue curves are in excellent agreement with the simulation results which shows that both the reflection and diffraction effects must be considered in the solution to obtain more accurate results. The values of the red curves are equivalent to the results of the infinite flange case.
	The values of the black and brown curves in Fig. 4(a) in the vicinity of the TE$_{10}$ cut-off frequency tend to infinity. This phenomenon can be explained by examining the complex propagation constant $\gamma_{uv}$ given in (A.3) for a lossless structure. At the cut-off frequency, $\gamma_{uv}$ becomes zero. Consequently, according to equation (17), NPR goes to infinity. However, when we account for the reflected fields, where the mode reflection coefficient is equal to -1 (i.e., no radiation), the numerator of the NPR equation also goes to zero. This is why NPR is being confined to a finite value, as shown in our results. Thus, ignoring the effect of the reflected fields due to the aperture may cause significant errors in the final results.
	
	The results of antenna factor for MoM and the measured values of an open-ended WR-90 waveguide mounted on an 11 cm $\times$ 10 cm flange are presented in \cite{ref8}. The antenna factor is defined as 
	\begin{equation}
		\textrm{AF} = \frac{E_{inc}}{V_{rec}}
	\end{equation}
	where $V_{rec}$ is the voltage received at the antenna port (or the spectrum analyzer terminal), and $E_{inc}=1$ V/m is assumed for the measured results presented in \cite{ref8}. Then the received power can be obtained from
	\begin{equation}
		P_{rec} = \frac{1}{2} V_{rec} I_{rec}^\ast
	\end{equation}
	where $I_{rec}$ is the induced current in the spectrum analyzer terminal. Since the frequency range covers only the dominant TE$_{10}$ mode bandwidth, we can write the magnitude of penetrated power of TE$_{10}$ mode as follows, by ignoring losses and impedance mismatch
	\begin{equation}
		|P^h_{10}| = \frac{|V_{rec}|^2}{2Z_{ref}}
	\end{equation}
	where $Z_{ref}$ is the measurement system reference impedance (here, 50$\Omega$). Using (17) and the concept of rectangular waveguide power defined in \cite{ref14}, we can calculate $P^h_{10}$ in terms of NPR$^h_{10}$ given as
	\begin{equation}
		|P^h_{10}| = \frac{(E_{inc}|\textrm{NPR}^h_{10}|)^2 a_1^2 ab \gamma_{10}}{\eta_0 k_0^3}.
	\end{equation}
	Using (18), and by combining (20) and (21), we can simply convert the antenna factor data to the magnitude of NPR$^h_{10}$.
	
	\begin{figure}[!t]
		\centering
		\includegraphics[width=3.3in]{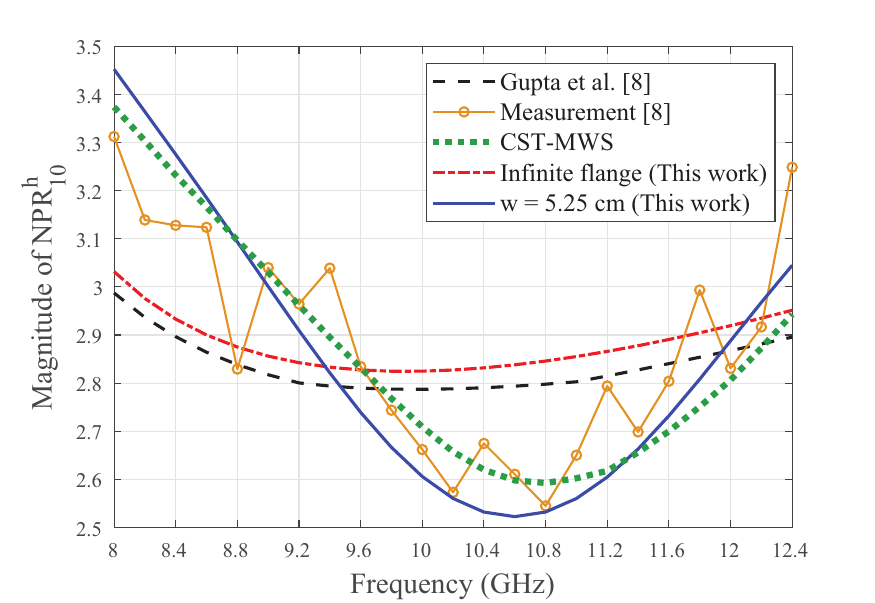}
		\caption{Magnitude of NPR$^h_{10}$ in terms of frequency changes for 11cm $\times$ 10cm flange.}
	\end{figure}
	In Fig. 5, the converted AF results presented in \cite{ref8} are shown. Also, the simulation results for the same flange, and the results of our method for the cases of $w = 5.25$ cm and infinite flange are included. It can be seen that the impact of the finite flange is not considered in those MoM results of \cite{ref8}, and a significant error occurs compared with the simulation results. However, our results for $w = 5.25$ cm are in good agreement with the simulation and measurement.

	\subsection{The Evaluation of NPR$^h_{10}$ in Terms of Flange side Changes}
	According to Appendix A and Appendix B, the diffraction coefficient and the radiated fields to the diffraction point (i.e., $\bar{E}^i(Q_d)$) depend on the length of the flange side. Thus, in Fig. 6, we show the magnitudes and phases of NPR$^h_{10}$ in terms of the flange side changes for three frequencies chosen to evaluate both propagating and evanescent cases of TE$_{10}$ penetrated fields. 
	\begin{figure}[!t]
		\centering
		\includegraphics[width=3.3in]{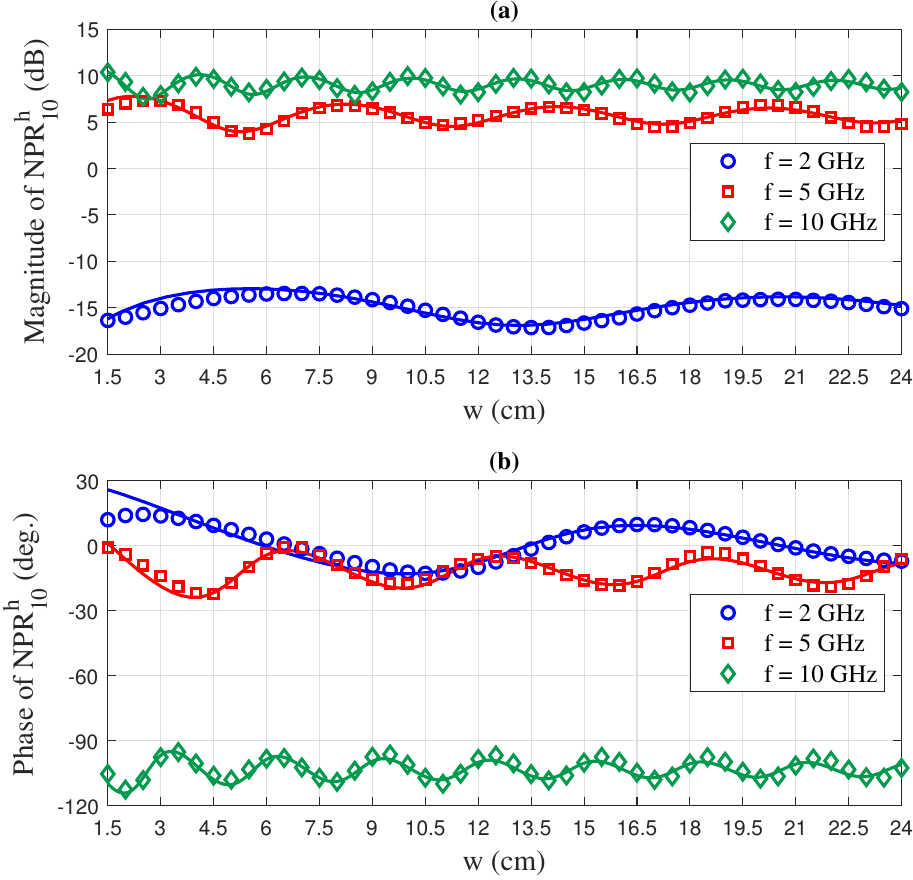}
		\caption{(a) Magnitude and (b) Phase of NPR$^h_{10}$ in terms of flange side length changes (solid curves: analytical, markers: simulation).}
	\end{figure}
	It can be seen that our results are in excellent agreement with the simulation results when $k.w > 4$. This is likely due to our assumption that $\bar{E}_a$ is radiated from the center of the aperture and $\bar{E}^i(Q_d)$ is in the far-field region at the diffraction point. However, when $k.w < 4$, our results are still in a similar trend to the simulations, and a maximum 1.2 dB difference is observed.
	
	Based on the far-field assumption, $\bar{E}^i(Q_d)$ is inversely proportional to $w$. According to (B.5) and (B.2), the spatial attenuation factor is directly proportional to $\sqrt{w}$, and the diffraction coefficient is no longer related to $w$ when $\theta = 0$ and $k.w > 10$. Therefore, using (15) under these conditions lead us to the decreasing rate of $1/\sqrt{w}$ for the magnitudes of the diffracted fields. This shows why the variations of the results in Fig. 6 decrease slowly when $w$ increases. Therefore, the flange must be very large to be approximated with an infinite flange.

	\subsection{The Evaluation of NPR$^h_{10}$ in Oblique Plane Wave Incidence}
	The previous examples evaluated the geometry of Fig. 6 when the plane wave is polarized in $\hat{y}$ direction. However, the proposed method can calculate the penetrated field for different angles of incidence. In Fig. 2, when the waveguide operates in TE$_{10}$ mode, it can be easily found from the radiated fields that the E-plane is the plane of $\phi = \pi/2$, and the H-plane is the plane of $\phi = 0$.
	\begin{figure}[!t]
		\centering
		\includegraphics[width=3.25in]{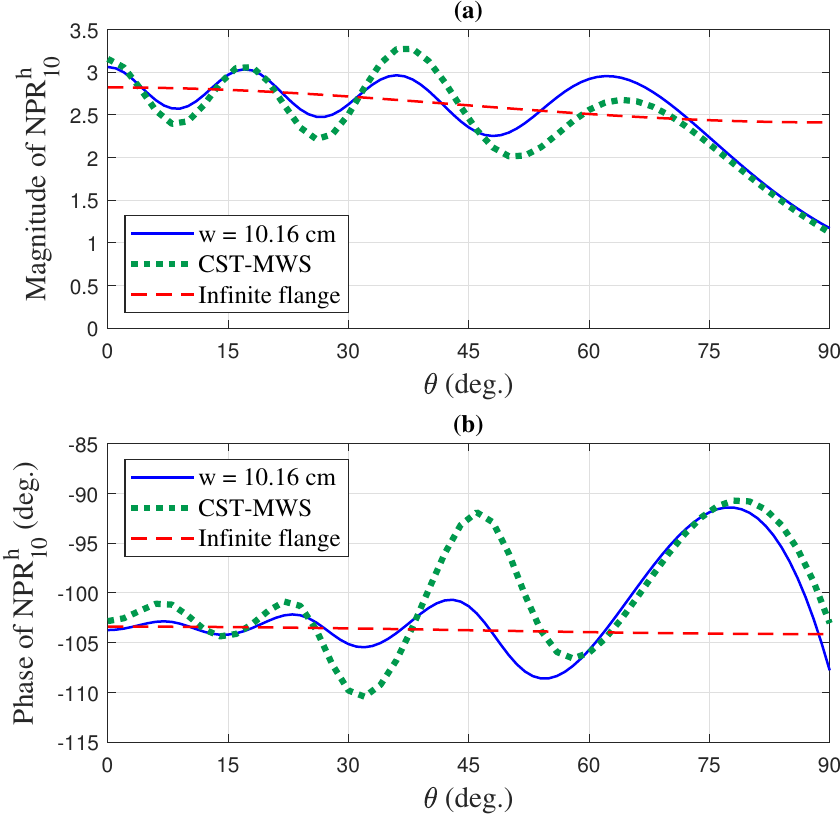}
		\caption{(a) Magnitude and (b) Phase of NPR$^h_{10}$ in E-plane ($\phi = \pi/2$ and $\psi = 0$) in terms of $\theta$ changes at $f = 10$ GHz.}
	\end{figure}
	Fig. 7 and Fig. 8 show the values of NPR$^h_{10}$ in terms of $\theta$ changes in E-plane and H-plane, respectively. Our results are in satisfactory agreement with the simulations results, and the observable differences are due to the approximate method of the diffraction calculation.
	\begin{figure}[!t]
		\centering
		\includegraphics[width=3.25in]{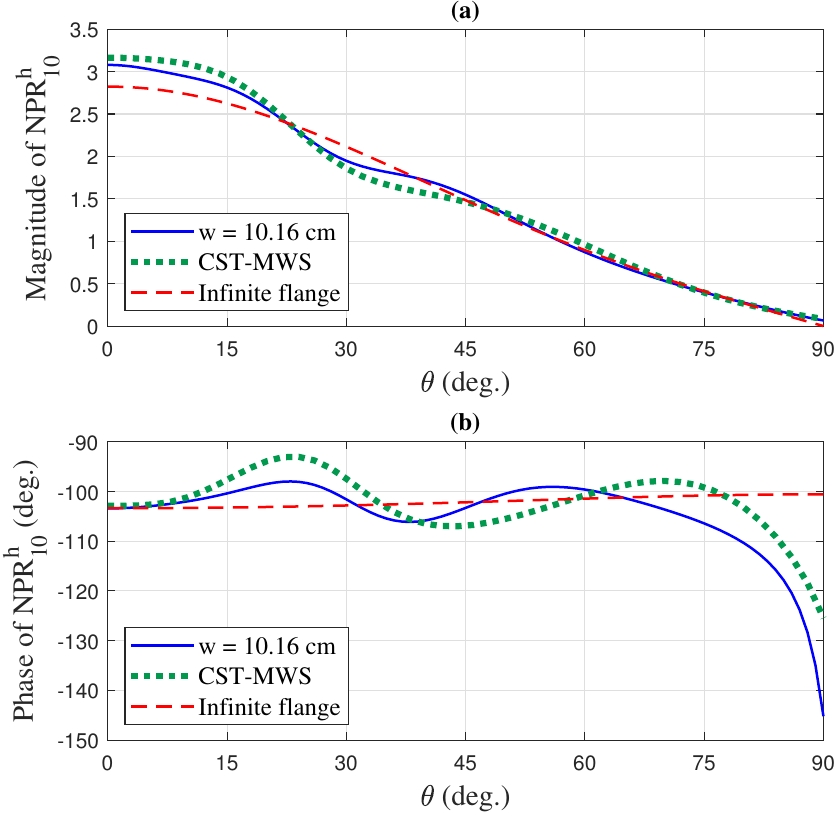}
		\caption{(a) Magnitude and (b) Phase of NPR$^h_{10}$ in H-plane ($\phi = 0$ and $\psi = \pi/2$) in terms of $\theta$ changes at $f = 10$ GHz.} 
	\end{figure}
	However, it is apparent that the calculated diffracted fields improve the results significantly compared with the infinite flange results.
	The slight difference between the analytical results in Fig. 7 and Fig. 8 at $\theta = 0$ is due to the diffraction calculation method. The results of Fig. 7 are obtained using first-order diffracted fields where the effect of other diffraction methods is insignificant. However, the results of Fig. 8 are obtained by the equivalent current contribution method and the second-order diffracted fields, because the first-order diffraction in H-plane is zero \cite{ref15}. Also, when $\theta$ increases, effect of the second-order diffracted fields increases too. This prevents the fields vanishing at $\theta = 90$\textdegree\ (grazing incidence), as shown in Fig. 8.

	\subsection{Higher-Order Modes Evaluation}
	\begin{figure}[!t]
		\centering
		\includegraphics[width=3.25in]{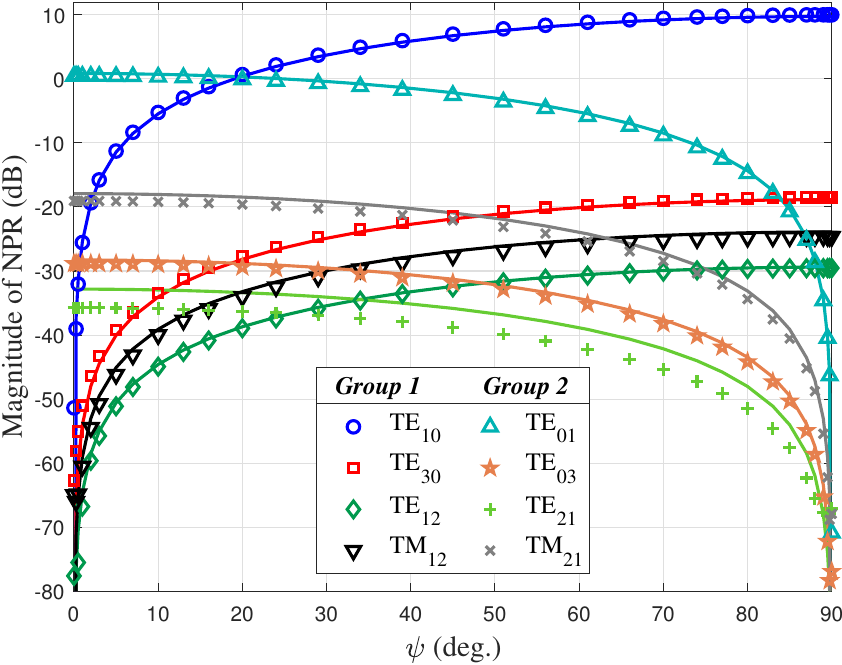}
		\caption{Magnitude of the NPR of the higher-order modes in terms of $\psi$ changes for $\phi = 0$ and $f = 10$ GHz (solid curves: analytical, markers: simulation).}
	\end{figure}
	Our solution given in (17) can be used to calculate the penetrated fields of the rectangular waveguide modes separately. We calculated only NPR$^h_{10}$ in the previous examples. However, a similar procedure is applied to calculate the penetrated fields of the higher-order modes. Generally, the rectangular waveguide modes are separated into four groups of coupled modes in terms of their indices being even or odd \cite{ref16}. Consequently, we can define Group 1 of modes in which their first index is odd and their second index is even (e.g., TE$_{10}$, TE$_{30}$, TE$_{12}$, TM$_{12}$, ...), and Group 2 of modes in which their first index is even and their second index is odd (e.g., TE$_{01}$, TE$_{03}$, TE$_{21}$, TM$_{21}$, ...). As a result of \cite{ref16}, $\bar{E}^{ap,r}$ consists of the modes that are only coupled with the mode of $\bar{E}^{ap,i}$. If we ignore the reflection effect (i.e., $\bar{E}^{ap,r} = 0$), it can be shown that when $\theta = 0$, only the sources of TE$_{m0}$ modes ($m = 2n - 1$ when $n\in \mathbb{N}$) radiate non-zero $\bar{E}^u$ in $\hat{y}$ direction. The same is true for TE$_{0m}$ modes in $\hat{x}$ direction. Now, if we include the reflection effect, the other modes in Group 1 and Group 2 can radiate $\bar{E}^u$ in those directions. Based on (17), the same is true for the case of penetration. Fig. 9 shows the magnitudes of NPR for four modes of the two groups in terms of $\psi$ changes when $\theta = 0$ and $\phi = 0$. Since $\theta = 0$, the modes in Group 1 and Group 2 are only the modes with a significant magnitude of NPR when $\psi$ tends to 90\textdegree\ and 0\textdegree, respectively. The results of the two other groups (i.e., the modes with both even or both odd indices) are not shown since they are too small which shows the insignificant diffraction effect for those modes.

	\subsection{The Evaluation of NPR$^h_{10}$ in Terms of the Waveguide Dimensions Changes}
	\begin{figure}[!t]
		\centering
		\includegraphics[width=3.3in]{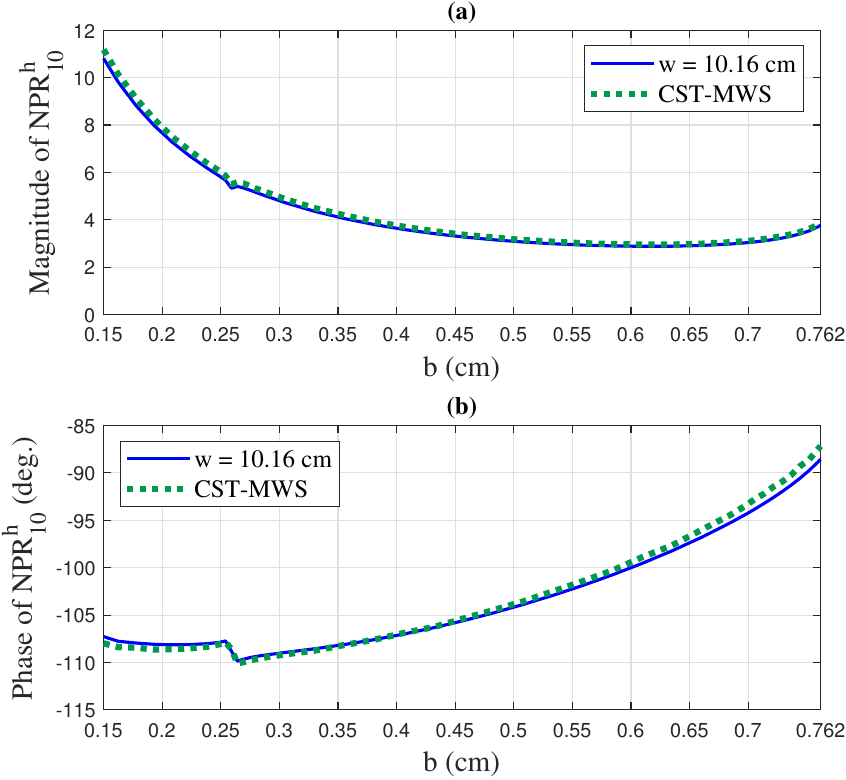}
		\caption{(a) Magnitude and (b) Phase of the NPR$^h_{10}$ in terms of $b$ changes for $a = 0.58/b$ cm at f = 10 GHz.}
	\end{figure}
	The problem of optimizing aperture dimensions to minimize (or maximize) the penetrated fields through the aperture plays an essential role in the practical case. We intend to evaluate the TE$_{10}$ mode penetrated field when the aperture surface remains constant so that $ab = 0.58$ cm$^2$ which is similar to the WR-90 waveguide. Fig. 10 shows the values of NPR$^h_{10}$ in terms of $b$ changes. When $b$ tends to smaller values, the aperture is more analogous to a narrow rectangular slot, and Fig. 10 (a) shows that in this case, the magnitude of NPR$^h_{10}$ increases significantly. Also, a small dip is observed at $b = 0.259$ cm. This dip is related to the coupling between TE$_{30}$ and TE$_{10}$ modes discussed in \cite{ref16} because for $b\simeq 0.259$ cm, we have $a\simeq 2.242$ cm which leads to the TE$_{30}$ cut-off frequency of around 10 GHz.
	
	\begin{figure}[!t]
		\centering
		\includegraphics[width=3.1in]{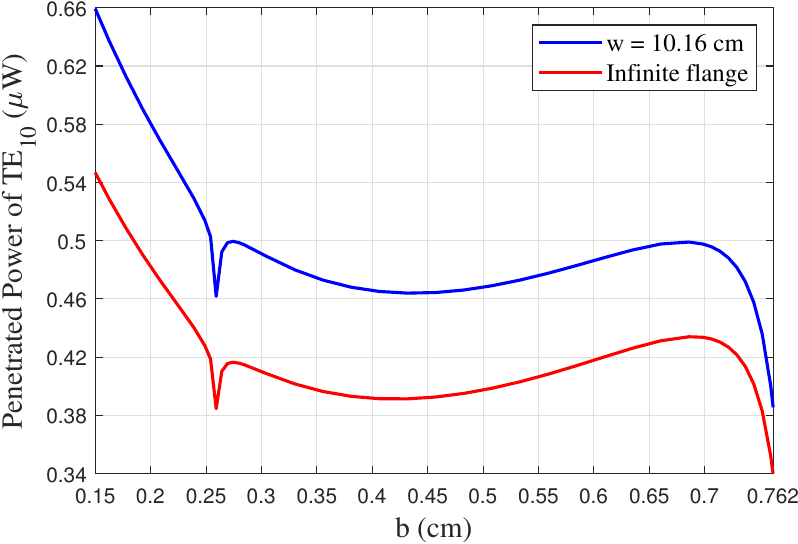}
		\caption{The penetrated power of TE$_{10}$ dominant mode in terms of $b$ changes for $a = 0.58/b$ cm at f = 10 GHz.}
	\end{figure}
	Fig. 11 shows the penetrated power of the dominant TE$_{10}$ mode through the aperture when the electric field amplitude of the incident plane wave is 1 V/m. The penetrated TE$_{10}$ mode is propagating for all values of $b$ in Fig. 11. To show the effect of diffraction on the penetrated power, we calculate the power of TE$_{10}$ mode using (21) by including and not including the diffracted fields in solution. In both cases, the maximum and minimum penetrated power occur at the boundaries of $b$ changes (minimum for square aperture). Also, there are four local extrema at $b= 0.259$, $0.275$, $0.432$, and $0.686$ cm. The first two extrema are caused by being near the TE$_{30}$ cut-off frequency as mentioned before. By comparing the results shown in Fig. 11, the values of TE$_{10}$ penetrated power for the finite flange ($w=10.16$ cm) is on average about 18\% greater than the values of the infinite flange.

	\section{Performance Benchmark}
	As previously discussed, a major advantages of the analytical solution over other methods (e.g., numerical or experimental) is its ability to reduce the calculation time significantly. To this end, we have evaluated our proposed geometry in Fig. 2 and its CST model in Fig. 3 with the primary specifications (discussed earlier in Section III) at frequencies of 5 GHz and 10 GHz. We measured the runtimes for the same problems using the 3D full-wave simulator and our MATLAB-based program and the results are presented in Table I. Both programs use the same computer configured with a 13\textsuperscript{th} Gen. Intel\textregistered\ Core\texttrademark\ i9 processor and 64 GB RAM. The results are reported in Table I. It is evident that our analytical method is more than 60 times faster than the conventional 3D full-wave numerical method. This demonstrates the superiority of our analytical model over other 3D full-wave numerical approaches. It should be noted that our code could be faster by implementing a low-level programming language like C.      
	
	\begin{table}[!h]
		\caption{Measured Runtime for Obtaining NPR$^h_{10}$ for the Proposed Geometry with its Primary Specifications}
		\centering
		\resizebox{230pt}{20pt}{
			\begin{threeparttable}
				\begin{tabular}{|c|c|c|}
					\hline
					Operating frequency& f = 5 GHz & f = 10 GHz \\
					\hline 
					Our MATLAB-based program& 1.95 s  & 1.97 s\\ 
					\hline 
					CST 3D full-wave simulator& 119.2 s & 126.9 s\\
					\hline
				\end{tabular}
		\end{threeparttable}}
	\end{table}
	
	Our MATLAB-based program consists of two stages to calculate NPR$^h_{10}$. In the first stage, we calculate the first column of the modal scattering matrix (MSM) that satisfies condition set $A$, as discussed in \cite{ref16}. In the second stage, we use the MSM results and the proposed GTD approach to accurately calculate the radiated field, allowing us to directly compute (17). Table II shows the detailed runtime of our MATLAB-based program. It is evident that most computational time is consumed in the first stage. This indicates that, without altering the aperture characteristics or the operating frequency, we can achieve results for other parameter changes much faster.
	
	\begin{table}[!h]
		\caption{Detailed MATLAB Code Runtime for Obtaining NPR$^h_{10}$ for the Proposed Geometry with its Primary Specifications}
		\centering
		\resizebox{230pt}{20pt}{
			\begin{threeparttable}
				\begin{tabular}{|c|c|c|}
					\hline
					& code's first stage & code's second stage \\
					\hline 
					f = 5 GHz & 1.86 s  & 0.08 s\\ 
					\hline 
					f = 10 GHz & 1.87 s & 0.09 s\\
					\hline
				\end{tabular}
		\end{threeparttable}}
	\end{table}
	
	\section{Conclusion}
	A modal analysis based on the reciprocity theorem is presented in this paper to calculate the penetrated fields into an open-ended rectangular waveguide mounted on a finite flange due to a plane wave incidence. The proposed method relates the penetrated fields of each waveguide mode to the far-field radiated components of that mode. Thus, the accuracy of the final results depends on the accuracy of the calculated radiated field components. The effects of both the reflected fields due to the waveguide aperture and the diffracted fields due to the flange edges must be considered in calculating the radiated field components. We have evaluated the geometry of a rectangular waveguide mounted on a thick square flange for different operating frequencies, flange sizes, plane wave incidence angles and polarization directions, and waveguide dimensions. All evaluated examples indicate an excellent agreement between our analytical results and simulations.
	
	Ignoring the effect of the reflected fields due to the wave-guide aperture yields an infinite magnitude of the penetrated field in the vicinity of the mode cut-off frequency. The variations in the penetrated fields decrease slowly when the flange side length increases because the magnitudes of the diffracted fields decrease with the rate of $1/\sqrt{w}$. Thus, the infinite flange approximation should be applied to a very large flange. Our results demonstrate that the penetrated fields are not zero at grazing incidences of the plane wave. We have shown that for a constant aperture surface, by changing its dimensions correspondingly, some perturbations in penetrated field results may happen due to the coupling between the modes at their cut-off frequencies. Also, we have shown that including the diffraction effect in the solution significantly improves the results of the dominant mode penetrated power. Moreover, it was shown that a major superiority of our analytical approach over the numerical approach is its much shorter calculation time which is reduced by a factor of 60.

	\appendices
	\numberwithin{equation}{section}
	\section{Far-field Radiation From a Rectangular Waveguide Mounted on an Infinite Flange}
	The solution for the reflected fields from the aperture of an open-ended rectangular waveguide for any incident waveguide mode is presented in \cite{ref16}. We use this approach to obtain the expressions for the far-field radiation of each waveguide mode from the aperture mounted on an infinite flange (i.e., $\bar{E}^u$).
	
	The electromagnetic solutions for the waveguide modes can be derived from electric and magnetic Hertz vector as
	\begin{multline}
		\bar{\Pi}^{h} (x,y,z) =\\
		\sum_{\substack{m,n=0\\m=n\neq0}}^{\infty} \frac{A_{mn}^{h}}{k_0^2 \eta_0} \cos{a_m (x+a)} \cos{b_n (y+b)} e^{\pm\gamma_{mn}z} \hat{z}
	\end{multline}
	\begin{multline}
		\bar{\Pi}^{e} (x,y,z) =\\
		\sum_{m,n=1}^{\infty} \frac{A_{mn}^{e}}{k_0^2} \sin{a_m (x+a)} \sin{b_n (y+b)} e^{\pm\gamma_{mn}z} \hat{z}
	\end{multline}
	where $A_{mn}$ is the mode amplitude coefficient at $z = z_0$, $\hat{z}$ is the unit vector along the z-axis, $a_m = m\pi/(2a)$, and $b_n = n\pi/(2b)$. The negative and the positive sign in exponential term belong to the incident and the reflected modes, respectively, and $\gamma_{mn}$ can be written as
	\begin{equation}
		\gamma_{mn} = \sqrt{a_m^2 + b_n^2 - k_0^2}.
	\end{equation}
	Also, the electric and magnetic fields are obtained from
	\begin{equation}
		\bar{E} = \bar{\nabla} (\bar{\nabla} \cdotp \bar{\Pi}^{e}) + k_0^2\bar{\Pi}^{e} - j\omega \mu_0 \bar{\nabla}\times \bar{\Pi}^{h}
	\end{equation}  
	\begin{equation}
		\bar{H} = \bar{\nabla} (\bar{\nabla} \cdotp \bar{\Pi}^{h}) + k_0^2\bar{\Pi}^{h} + j\omega \varepsilon_0 \bar{\nabla}\times \bar{\Pi}^{e}
	\end{equation}
	Based on \cite{ref16}, we can calculate the amplitude coefficients of the reflected modes using the following expressions
	\begin{multline}
		\sum_{m,n=1}^{\infty} [a_m I_1(m,n,p,q) + b_n I_2(m,n,p,q)] \gamma_{mn} A_{mn}^{re} \\
		+\sum_{\substack{m,n=0\\m=n\neq0}}^{\infty} [b_n I_1(m,n,p,q) - a_m I_2(m,n,p,q)] jk_0 A_{mn}^{rh} \\
		-\sum_{u,v=1}^{\infty} [a_u I_1(u,v,p,q) + b_v I_2(u,v,p,q)-\frac{jk_0}{\gamma_{uv}}b_v \epsilon_v \delta] \gamma_{uv} A_{uv}^{ie} \\
		+\sum_{\substack{u,v=0\\u=v\neq0}}^{\infty} [b_v I_1(u,v,p,q) - a_u I_2(u,v,p,q) + \frac{\gamma_{uv}}{jk_0}a_u \epsilon_v \delta] jk_0 A_{uv}^{ih} \\
		= - ab[j k_0 b_q A_{pq}^{re} - \gamma_{pq} a_p \epsilon_q A_{pq}^{rh}]
	\end{multline}
	for $p = 1,2,3,...$ and $q = 0,1,2,...$, and
	\begin{multline}
		\sum_{m,n=1}^{\infty} [a_m I_3(m,n,p,q) + b_n I_4(m,n,p,q)]\gamma_{mn} A_{mn}^{re}\\
		+\sum_{\substack{m,n=0\\m=n\neq0}}^{\infty} [b_n I_3(m,n,p,q) - a_m I_4(m,n,p,q)]jk_0 A_{mn}^{rh}\\
		-\sum_{u,v=1}^{\infty} [a_u I_3(u,v,p,q) + b_v I_4(u,v,p,q)-\frac{jk_0}{\gamma_{uv}}a_u \epsilon_u \delta]\gamma_{uv} A_{uv}^{ie}\\
		+\sum_{\substack{u,v=0\\u=v\neq0}}^{\infty} [b_v I_3(u,v,p,q) - a_u I_4(u,v,p,q) - \frac{\gamma_{uv}}{jk_0}b_v \epsilon_u \delta]jk_0 A_{uv}^{ih}\\
		= - ab[j k_0 a_p A_{pq}^{re} + \gamma_{pq} b_q \epsilon_p A_{pq}^{rh}]
	\end{multline}
	for $p = 0,1,2,...$ and $q = 1,2,3,...$, where
	\begin{equation}
		{\epsilon_i} = \begin{cases}
			2,&{\text{if}}\ i=0 \\ 
			{1,}&{\text{otherwise.}} 
		\end{cases}
	\end{equation} 
	\begin{equation}
		{\delta} = \begin{cases}
			ab,&{\text{if}}\ u=p \text{ and } v=q \\ 
			{0,}&{\text{otherwise.}} 
		\end{cases}
	\end{equation}
	\begin{multline}
		I_1(m,n,p,q) = \frac{1}{4\pi^2}\int_{-\infty}^{\infty} \int_{-\infty}^{\infty} \frac{\xi\eta}{\zeta k_0} C_m^a(-\xi)S_n^b(-\eta)\\ \times S_p^a(\xi)C_q^b(\eta) d\eta d\xi
	\end{multline}
	\begin{multline}
		I_2(m,n,p,q) = \frac{1}{4\pi^2}\int_{-\infty}^{\infty} \int_{-\infty}^{\infty} \frac{(k_0^2 - \xi^2)}{\zeta k_0} S_m^a(-\xi)C_n^b(-\eta)\\ \times S_p^a(\xi)C_q^b(\eta) d\eta d\xi
	\end{multline}
	\begin{multline}
		I_3(m,n,p,q) = \frac{1}{4\pi^2}\int_{-\infty}^{\infty} \int_{-\infty}^{\infty} \frac{(k_0^2 -\eta^2)}{\zeta k_0} C_m^a(-\xi)S_n^b(-\eta)\\ \times C_p^a(\xi)S_q^b(\eta) d\eta d\xi
	\end{multline}
	\begin{equation}
		I_4(m,n,p,q) = I_1(p,q,m,n)
	\end{equation}
	\begin{equation}
		S_p^a(\xi) = \int_{-a}^{a} \sin{a_p (x+a)} e^{-j\xi x} dx
	\end{equation}
	\begin{equation}
		C_q^b(\eta) = \int_{-b}^{b} \cos{b_q (y+b)} e^{-j\eta y} dy
	\end{equation}
	\begin{equation}
		C_p^a(\xi) = \int_{-a}^{a} \cos{a_p (x+a)} e^{-j\xi x} dx
	\end{equation}
	\begin{equation}
		S_q^b(\eta) = \int_{-b}^{b} \sin{b_q (y+b)} e^{-j\eta y} dy
	\end{equation}
	
	Once the amplitude coefficients of all considered modes are obtained, the far-field radiation components for different modes with specific $pq$ indices can be calculated from the following expressions. These expression are obtained using (13) and the far-field radiation equations for rectangular aperture antennas presented in \cite{ref22}
	\begin{multline}
		E^{uh}_{\theta} = \frac{A^h_{pq}}{2\pi}(a_p \sin(\phi) C_q^b(\frac{-Y}{b}) S_p^a(\frac{-X}{a})\\
		- b_q \cos(\phi) S_q^b(\frac{-Y}{b}) C_p^a(\frac{-X}{a}))
	\end{multline}
	\begin{multline}
		E^{uh}_{\phi} = \frac{A^h_{pq}}{2\pi}(a_p \cos(\theta) \cos(\phi) C_q^b(\frac{-Y}{b}) S_p^a(\frac{-X}{a})\\
		+ b_q \cos(\theta) \sin(\phi) S_q^b(\frac{-Y}{b}) C_p^a(\frac{-X}{a}))
	\end{multline}	
	\begin{multline}
		E^{ue}_{\theta} = \frac{A^e_{pq} \gamma_{pq} }{j2\pi k_0}(b_q \sin(\phi) C_q^b(\frac{-Y}{b}) S_p^a(\frac{-X}{a})\\ + a_p \cos(\phi) S_q^b(\frac{-Y}{b}) C_p^a(\frac{-X}{a}))
	\end{multline}
	\begin{multline}
		E^{ue}_{\phi} = \frac{A^e_{pq} \gamma_{pq}}{j2\pi k_0}(b_q \cos(\theta) \cos(\phi) C_q^b(\frac{-Y}{b}) S_p^a(\frac{-X}{a})\\- a_p \cos(\theta) \sin(\phi) S_q^b(\frac{-Y}{b}) C_p^a(\frac{-X}{a}))
	\end{multline}
	where
	\begin{equation}
		X = k_0 a \sin(\theta)\cos(\phi)
	\end{equation}
	\begin{equation}
		Y = k_0 b \sin(\theta)\sin(\phi).
	\end{equation}

	\section{Diffraction From the Edges of a Thick Square Flange}
	
	\subsection{First-Order Diffraction}
	As mentioned in Section II-C, we can use (15) to calculate the diffracted fields. For the geometry of Fig. 2, the different terms in (15) are determined as follows. We assume that the incident field on the diffraction points (i.e., $\bar{E}^i(Q_{d1,d2,d3,d4})$) are radiated from the center of the aperture using far-field approximation. Thus, on one hand, we can use the results of $\bar{E}^u$ at the diffraction points for calculating $\bar{E}^i(Q_d)$. On the other hand, according to \cite{ref14} and \cite{ref15}, $s' = w$, and we have
	\begin{equation}
		{s} = \begin{cases}
			r,&{\text{for amplitude terms}} \\ 
			{r - w \sin(\theta),}&{\text{for phase terms at }Q_{d1,d2}}\\
			{r + w \sin(\theta),}&{\text{for phase terms at }Q_{d3,d4}} 
		\end{cases}
	\end{equation}
	
	The first-order dyadic diffraction coefficient for the presented geometry in Fig. 2 is simplified to
	\begin{multline}
		\bar{\bar{D}} = \pm D_s\hat{\phi_d}\hat{\phi} \pm D_h\hat{\theta_d}\hat{\theta}\\
		=\mp \frac{e^{-j\pi/4}}{\tau\sqrt{2k_0\pi}\sin\beta_0'}\{\cot(\frac{\pi + \vartheta}{2\tau})F(kwg^+)\\
		+\cot(\frac{\pi - \vartheta}{2\tau})F(kwg^-)\}\hat{\theta_d}\hat{\theta}
	\end{multline}
	where the upper sign belongs to the fields at $Q_{d1,d2}$ and the lower sign belongs to the fields at $Q_{d3,d4}$, $F(x)$ is the Fresnel integral, $\hat{\theta_d}$ and $\hat{\phi_d}$ are unit vectors along $\hat{\theta}$ and $\hat{\phi}$ at the diffraction points, respectively, and
	\begin{equation}
		{\vartheta} = \begin{cases}
			\pi/2 + \theta,&{\text{at }Q_{d1,d2}} \\ 
			{\pi/2 - \theta,}&{\text{for } 0\leq \theta \leq \pi/2\text{ at }Q_{d3,d4}}\\
			{5\pi/2 - \theta,}&{\text{for } \pi/2 < \theta \leq \pi\text{ at }Q_{d3,d4}} 
		\end{cases}
	\end{equation}
	\begin{equation}
		g^\pm = 1 + \cos(2\tau\pi N^\pm - \vartheta)
	\end{equation}
	where $N^+$ and $N^-$ are chosen from integer numbers so that the cosine arguments in (B.4) be the nearest values to $+\pi$ and $-\pi$, respectively. Also, for the spherical wavefront at the diffraction point we have
	\begin{equation}
		A(s',s) = \frac{\sqrt{w}}{r}.
	\end{equation}
	It must be noted that for the grazing incidence on the diffraction points, (15) must be multiplied by the factor of $1/2$. Then, $\bar{E}^d$ for first-order diffraction is the summation of (15) for the diffraction points.

	\subsection{Second-Order Diffraction}
	According to (15), when the incident field on the diffraction point goes to zero the first-order diffraction goes to zero too. However, the diffracted fields are not zero in practical \cite{ref14}. Therefore, we use the second-order (slope) diffraction expressions at the diffraction points in order to improve the accuracy of our result as follows
	\begin{equation}
		\bar{E}^d(s) = \frac{1}{jk_0}\left(\frac{\partial \bar{E}^i(Q_d)}{\partial n}\right)\cdot \left(\frac{\partial \bar{\bar{D_s}}}{\partial \phi'}\right) \sqrt{\frac{s'}{s(s'+s)}} e^{-jk_0s}
	\end{equation}
	where
	\begin{equation}
		\frac{\partial \bar{E}^i(Q_d)}{\partial n}=-\frac{1}{w}\frac{\partial \bar{E}^u}{\partial \theta}\arrowvert_{Q_d}
	\end{equation}
	\begin{multline}
		\frac{\partial \bar{\bar{D_s}}}{\partial \phi'} = \mp \frac{e^{-j\pi/4}}{2\tau^2\sqrt{2k_0\pi}\sin\beta_0'}\{\csc^2(\frac{\pi + \vartheta}{2\tau})F_s(kwg^+)\\
		-\csc^2(\frac{\pi - \vartheta}{2\tau})F_s(kwg^-)\}\hat{\phi_d}\hat{\phi}
	\end{multline}
	\begin{equation}
		F_s(X) = 2jX(1-F(X)).
	\end{equation}

	\subsection{Equivalent Current Contribution Method}
	Based on (B.2), the results of first-order diffraction do not impact on H-plane fields. Thus, we use the concept of equivalent current in diffraction \cite{ref14} which states that the diffracted fields from an edge can be generated by an equivalent current along the edge. Since the soft polarization diffraction coefficient is zero for the presented geometry of Fig. 2, only equivalent magnetic current can exist which can be obtained from
	\begin{equation}
		I^m_x = - \frac{\eta_0\sqrt{8\pi k_0}}{k_0} e^{-j\pi/4} H^u_x(Q_d) D_h.
	\end{equation}
	Therefore, the diffracted field due to this equivalent current can be obtained from
	\begin{equation}
		H^m_{\alpha} = j \frac{k_0 e^{-jk_0r}}{4\pi\eta_0r} \sin\alpha \int^{w}_{-w} I^m_x(x')e^{jk_0x'\cos\alpha}dx'
	\end{equation}
	which is numerically calculated. It should be noted that this field only impacts on H-plane fields. An vector analysis must be utilized to convert fields in $\hat{\alpha}$ direction to $\hat{\theta}$ and $\hat{\phi}$ directions, and it is not mentioned here for the sake of briefness.

	\section*{Acknowledgments}
	The authors gratefully acknowledge feedback from Prof. M. S. Majedi at the Ferdowsi University of Mashhad, Mashhad, Iran, and help from all members of EMC and Microwave Technology Research Lab at the Ferdowsi University of Mashhad.

\end{document}